\documentclass[aps,prl,twocolumn,superscriptaddress,amsfonts,amssymb,amsmath]{revtex4-2}

\usepackage{graphicx}   

\newcommand{\vWe}{\mbox{$\vec{\omega} _\text{emitt}$}}
\newcommand{\vWi}{\mbox{$\vec{\omega} _\text{imp}$}}
\newcommand{\Wi}{\mbox{$\omega _\text{imp}$}}
\newcommand{\We}{\mbox{$\omega _\text{emitt}$}}
\newcommand{\Wrms}{\mbox{$\omega ^\text{rms}_\text{imp}$}}

\begin{document}


\title{Transparent Spin Method for Spin Control of Hadron Beams in Colliders}


\author{Yu.N. Filatov}
\affiliation{Moscow Institute of Physics and Technology, Dolgoprudny, Moscow Region, Russia}
\author{A.M. Kondratenko}
\affiliation{Moscow Institute of Physics and Technology, Dolgoprudny, Moscow Region, Russia}
\affiliation{Science \& Technique Laboratory ``Zaryad'', Novosibirsk, Russia}
\author{M.A. Kondratenko}
\affiliation{Moscow Institute of Physics and Technology, Dolgoprudny, Moscow Region, Russia}
\affiliation{Science \& Technique Laboratory ``Zaryad'', Novosibirsk, Russia}
\author{Ya.S. Derbenev}
\affiliation{Thomas Jefferson National Accelerator Facility, Newport News, VA, USA}
\author{V.S. Morozov}
\affiliation{Thomas Jefferson National Accelerator Facility, Newport News, VA, USA}



\begin{abstract}
We present a concept for control of the ion polarization, called a transparent spin method.
The spin transparency is achieved by designing such a synchrotron structure that the net spin rotation angle in one particle turn is zero. The polarization direction of any ions including deuterons can be efficiently controlled using weak quasi-static fields. These fields allow for dynamic adjustment of the polarization direction during an experiment. The main features of the Transparent Spin method are illustrated in a figure-8 collider. The results are relevant to the Electron-Ion Collider considered in the US, the ion-ion collider NICA constructed in Russia, and a polarized Electron-ion collider planned in China.
\end{abstract}


\maketitle


\textit{Introduction.}~---
Polarized beam experiments have been and remain a crucial tool in understanding particle and nuclear structure and reactions from the first principles \cite{b:principles}. In particular, polarized light ion ($p$, $d$, ${}^3He$) and electron beams are necessary for the successful operation of a proposed high-luminosity polarized \textit{Electron-Ion Collider} (EIC) that is currently under active design \cite{b:EIC_MEIC,b:EIC_eRHIC,b:EIC_China}. Technologies for production of polarized light ions already exist or will be operational in the near future at several colliders worldwide~\cite{b:IonSource}. 

After a polarized beam is generated by a source, the first task is to preserve the beam's polarization in the process of its acceleration to the final energy.  
In general, this task has complete and efficient solutions based on using the \textit{Siberian Snake}~\cite{b:AccSnake1,b:AccSnake2,b:SYLee} and \textit{figure-8 orbit} techniques~\cite{b:AccFig8}. Both techniques eliminate depolarizing \textit{spin resonances}~\cite{b:SYLee} by making the spin tune energy independent. 
Presently, polarization preservation using helical snakes~\cite{b:RHICSnakes} has been successfully demonstrated up to high energies in a proton-proton collider, RHIC.
Other techniques for suppressing depolarization during acceleration have also been proposed, for example, by employing high super-periodicity in the design of a synchrotron~\cite{b:HighPeriodicity}. 

The next task is to maintain the polarization during a long-term experimental run when the main difficulties with preserving the polarization are associated with higher-order nonlinear spin-orbit resonances~\cite{b:LongTermTrack1,b:LongTermTrack2,b:LongTermTrack3,b:LongTermTrack4}. As RHIC experience shows~\cite{b:pol_RHIC}, selection of optimal betatron tunes and elimination of the energy dependence of the spin tune as in the TS mode allow for sustaining the polarization for many hours.  

A typical task of a collider setup is adjustment of the required polarization direction at the interaction point (IP). In RHIC, \textit{fixed} longitudinal polarization at a specific IP is set using a pair of \textit{spin rotators} turning the spins by $\pm 90^\circ$ \cite{b:RHICSnakes}. Another task is change of the polarization direction during an experiment and, in particular, frequent flips of the polarization to minimize systematic errors. Experimentally verified spin-flipping schemes are based on adiabatically sweeping an RF magnet's frequency through an induced spin resonance \cite{b:RFflip}. This technique is used in RHIC for spin flipping with an  efficiency of 97\% in the energy range of 24 to 255~GeV~\cite{b:RFflipRHIC}. However, every single crossing of an RF resonance causes some polarization loss that limits the admissible number of spin flips during an experiment.

This paper presents a new method for ion polarization control called a \textit{Transparent Spin} (TS) technique. This technique allows one to first preserve the polarization during acceleration and maintain it in the collider mode. It then enables flexible and efficient manipulation of the polarization direction of any particle species. The polarization can be adjusted to any direction at any orbital location during the whole time of an experiment. Such an adjustment causes practically no polarization~loss.

The ideas of the TS method were first formulated in the design process of the figure-8 booster and collider synchrotrons of the JLEIC project \cite{b:baseJLEIC}. They were further developed and applied in the design of the racetrack rings of the NICA hadron collider project \cite{b:NICA1}. 

\textit{Transparent spin concept.}~---
According to the basic theorem about the spin motion of a particle moving along an arbitrary periodic closed orbit~\cite{b:axisN}, there always exists a periodic spin axis $\vec{n}(z)=\vec{n}(z+C)$ such that the spin motion can be in general represented as precession about $\vec{n}(z)$ with a spin tune $\nu$. Here $z$ is the distance along the closed orbit, $C$ is the orbit circumference, and $2\pi\nu$ is the phase advance of the spin precession per particle turn. The $\vec{n}(z)$ axis is unique if the spin precession frequency is not a harmonic of the particle circulation frequency. Due to the spin tune spread associated with beam emittances the resulting beam polarization is established along $\vec{n}$~\cite{b:AlongAxisN}.

The TS method is based on designing such a magnetic structure of a synchrotron that the net effect of the synchrotron elements on the spin for motion along the design closed orbit is compensated over a single particle turn. The magnetic lattice of the synchrotron becomes ``transparent'' to the spin. A natural example of such a structure is a flat figure-8 ring. In ``transparent'' structures, the spin motion becomes degenerate: any spin direction at any orbital location repeats every particle turn. This means that the particles are in a \textit{TS resonance} with $\nu=0$. The spin motion in such a situation is highly sensitive to small perturbations of the magnetic fields along the orbit. At the same time, this sensitivity allows one to implement a simple and efficient spin control system using a \textit{spin navigator} (SN). An SN is a flexible device consisting of elements with weak \textit{constant} or \textit{quasi-stationary} magnetic fields rotating the spins about a desirable direction $\vec{n}_N$ by a small angle $2\pi\nu_N$. Such a navigator has practically no effect on the orbital beam dynamics~\cite{b:SmallSol,b:smallField}.

In an ideal synchrotron lattice, the stable polarization axis $\vec{n}$ in the straight housing the SN coincides with the SN axis $\vec{n}_N$ and the resulting spin tune $\nu$ equals $\nu_N$. Clearly, to control the spin, one must use a sufficiently strong SN to dominate over the effect of the small perturbative fields around a particle's trajectory. There are two sources of these fields: lattice imperfections and focusing magnetic and bunching RF fields experienced by particles during their free transverse and longitudinal (betatron and synchrotron) oscillations.

Let us illustrate the TS method using a figure-8 orbit as an example.

\textit{Spin motion in a figure-8 synchrotron.}~---
In an ideal figure-8 synchrotron, rotation of the spins in one arc is compensated by an opposite rotation in the other arc. When a particle is moving on a flat design orbit, the spin tune is zero for any particle energy, so the spin motion is degenerate.

When a particle's trajectory deviates from the design orbit, the particle experiences perturbing magnetic fields. The effect of these fields on the spin in a large number of particle turns can be expressed in terms of the \textit{TS resonance field} $\vec{\omega}$ (TS-field). In the absence of SNs, the direction of $\vec{\omega}$ defines the direction of the stable spin axis $\vec{n}=(\vec{\omega}/\omega)$ while its absolute value, or the \textit{TS resonance strength}, $\omega=|\vec{\omega}|$ defines the particle's spin tune: $\nu=\omega$, i.e. the spin completes a full rotation about the $\vec{n}$ axis in $1/\omega$ particle turns around the orbit~\cite{b:EIC_MEIC}.
The TS resonance spin field is a 3D extension of the conventional 2D resonance strength concept for cases when the spin motion is degenerate. Conventional codes, such as DEPOL~\cite{b:DEPOL}, are valid for synchrotrons with a distinct polarization direction and calculate the spin field component transverse to this direction.


To set the required polarization direction, an SN is inserted into the synchrotron lattice. For example, a weak solenoid inserted into an experimental straight stabilizes the longitudinal polarization direction in that straight.


\textit{Strength of the TS resonance.}~---
The TS-field $\vec{\omega}$ consists of two parts: the coherent field \vWi{} associated with the closed orbit distortion caused by lattice imperfections and the incoherent field \vWe{} associated with the beam emittances. 

With \textit{fixed} element alignment errors, the \vWi{} field of a synchrotron is constant and does not change during an experiment. Since alignment errors are random, our analysis uses an rms strength \Wrms{} obtained statistically assuming independent distributions of element misalignments~\cite{b:EIC_MEIC,b:statModel}. The magnitude of \vWe{} is non-zero only in the second order of the particle's oscillation amplitudes and is proportional to the beam emittances. In practice, \Wi{} significantly exceeds \We{}~\cite{b:EIC_MEIC}. 

\begin{figure}[b]
 \includegraphics[width=\columnwidth]{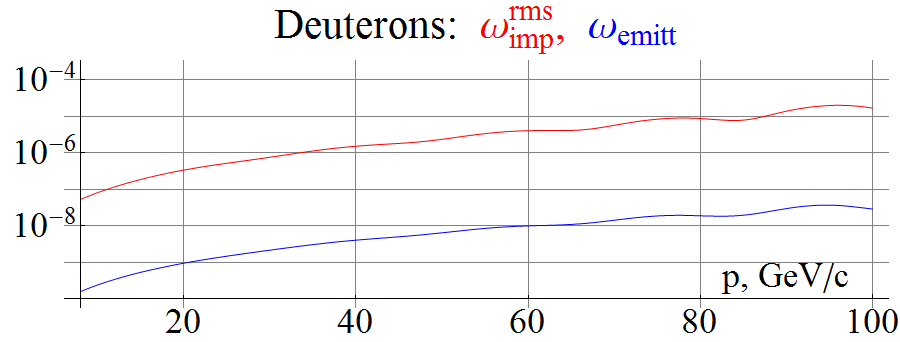}%
 \\[2mm]
 \includegraphics[width=\columnwidth]{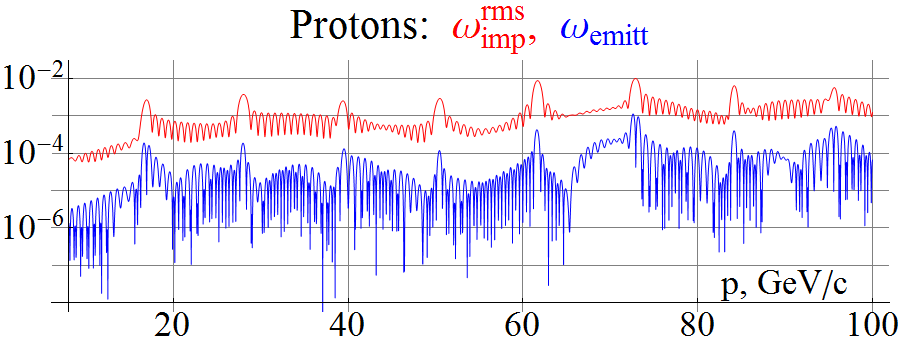}
 \caption{\label{f:W_JLEIC} TS resonance strengths for deuterons and protons in JLEIC versus the beam momentum.}
 \end{figure}

Figure~\ref{f:W_JLEIC} shows an analytic calculation of \Wrms{} and \We{} for deuterons and protons as functions of the beam momentum for JLEIC's figure-8 collider ring \cite{b:AccJLEIC}. 
\Wrms{} is calculated assuming random uncorrelated transverse shifts of all quadrupoles giving an rms closed orbit excursion of 100~$\mu$m. The calculation of \We{} assumes normalized transverse emittances of 1 mm$\cdot$mrad. 

The deuteron \Wrms{}  grows with momentum but does not exceed $3\cdot 10^{-5}$. The proton \Wrms{} is $ 10^{-4}$ -- $10^{-3}$ in the whole momentum range with exception of narrow interference peaks where the spin effects of the arc magnets add up coherently. In both cases, \We{} is about two orders of magnitude lower than \Wrms{} in the whole momentum range.

\textit{Polarization control condition.}~---
In the presence of a navigator, the particle spins are precessing about an effective spin field $\vec{h}$ consisting of the navigator and TS fields~\cite{b:axisN}:
\begin{equation} 
\vec{h}(z)=\nu_N \,\vec{n}_N(z)+\vec{\omega}(z)\,.
\end{equation}
At the SN location, $\vec{n}_N(z)$ coincides with the SN's spin rotation axis. In the rest of the ring, $\vec{n}_N(z)$ evolves according to the Thomas-BMT equation. 

To stabilize and control the polarization, the navigator strength must significantly exceed the TS resonance strength (the control condition)
\begin{equation}
\nu_N \gg \omega \,.
\end{equation}

Polarization then points along the navigator field direction $\vec{n}_N$. The angle of uncontrolled deviation of the polarization from $\vec{n}_N$ is of the order of $\omega/\nu_N$.
 
In the above example, depending on the beam momentum, the SN strength must be large compared to \mbox{$10^{-7}$--$10^{-5}$} for deuterons and $10^{-4}$--$10^{-2}$ for protons to control their polarization.

\textit{Acceleration in TS mode.}~---
Note that $\vec{h}$ is a function of energy. Typically, during acceleration, $\vec{h}$ changes adiabatically slowly, i.e. in a characteristic time of the field change, the spin makes a large number of turns about the field. In that case, the spin initially oriented along $\vec{h}$ follows it during the whole acceleration process and polarization is preserved. We showed analytically and numerically that, in the presence of an appropriate spin navigator, effects of imperfections, coupling, betatron and synchrotron oscillations can be neglected~\cite{b:EIC_MEIC}.

\begin{figure}[b]
 \includegraphics[width=\columnwidth]{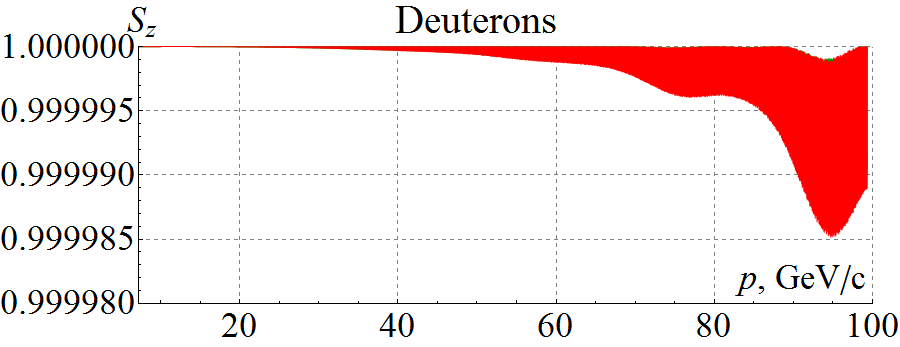}%
 \caption{\label{f:Sz_deut} Longitudinal spin component of a deuteron as a function of the beam momentum during acceleration in JLEIC.}
 \end{figure}

Figure~\ref{f:Sz_deut} shows the longitudinal spin component of a deuteron during acceleration in JLEIC \cite{b:AccJLEIC}. The simulation is done using a spin tracking code Zgoubi \cite{b:Zgoubi}. The simulation parameters are chosen based on a prediction of the same model that is used to analytically calculate the TS-resonance strength in Fig.~\ref{f:W_JLEIC}. The field ramp rate is 3~T/min. The longitudinal polarization direction is stabilized by a solenoid with a maximum field integral of about 7~T$\cdot$m. Such a solenoid provides a navigator strength of $\nu_N=3\cdot 10^{-3}$ and the control condition is satisfied with a large margin. The change in the longitudinal spin component does not exceed a value of $2\cdot 10^{-5}$ as shown in Fig.~\ref{f:Sz_deut}.

\begin{figure}[t]
 \includegraphics[width=\columnwidth]{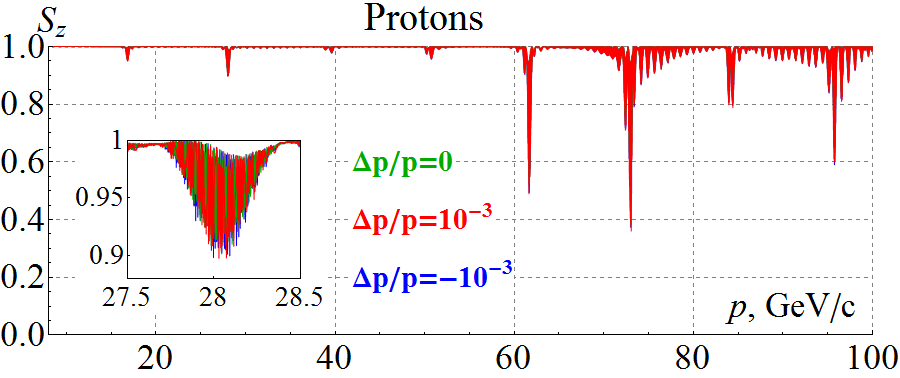}%
 \caption{\label{f:Sz_prot} Longitudinal spin component during acceleration of three protons in JLEIC.}
 \end{figure}

When accelerating protons in the same lattice, the same solenoid provides an SN strength of \mbox{$\nu_N= 10^{-2}$}. Figure~\ref{f:Sz_prot} shows the longitudinal spin components of three protons with $\Delta p/p=0$ (green line), $\Delta p/p=10^{-3}$ (red line) and $\Delta p/p=-10^{-3}$ (blue line). The graphs of the longitudinal spin components practically do not differ from each other (the red line covers up the blue and green lines), i.e. the synchrotron energy modulation has no noticeable effect on the ion spin motion.

The control condition in Fig.~\ref{f:Sz_prot} is met almost everywhere except for narrow energy regions of the interference peaks (see Fig.~\ref{f:W_JLEIC}) where the navigator strength is comparable to \Wrms{}. However, crossing of these peaks during acceleration does not depolarize the beam. As illustrated in Fig.~\ref{f:Sz_prot}, they only cause coherent deviation of the spins from the longitudinal direction by angles determined by both the navigator and
the coherent part of the  
TS-resonance strengths. The beam polarization restores its longitudinal direction in places where the control condition is again satisfied. 
This indicates that, in this example, the change in spin occurs adiabatically during acceleration.
 These tracking results agree well with the analytic calculation of the TS resonance strength in Fig.~\ref{f:W_JLEIC}. 

This example suggests that if the goal is to simply preserve the polarization degree, the spin control condition can be relaxed to a weaker one of $\nu_N \gg \We$. Direction control can be restored by accounting for \Wi{} in the SN setting. \Wi{} can be measured experimentally.

\textit{Spin navigators.}~---
Spin navigators can be technically realized in different ways using longitudinal and transverse fields~\cite{b:baseJLEIC,b:smallField}. As an example, let us consider a spin navigator design based on weak solenoids, which have no effect on the closed orbit. 

\begin{figure}[ht]
 \includegraphics[width=0.9\columnwidth]{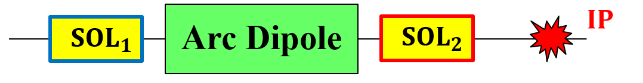}%
 \caption{\label{f:SN_2sol} Schematic of a spin navigator for control of the ion polarization in the collider's plane using small solenoids.}
 \end{figure}

A single solenoid stabilizes the longitudinal polarization direction at its location. One can set any polarization direction in the horizontal plane ($xz$) at the IP by introducing a second solenoid into the collider's lattice. 
Figure~\ref{f:SN_2sol} shows a schematic of such a spin navigator where the two solenoids 
are separated by an arc dipole~\cite{b:SmallSol}. 

One can similarly design a 3D spin navigator for control of all three polarization direction components~\cite{b:baseJLEIC}.

\textit{TS collider features.}~---
Let us briefly formulate some of the new capabilities that become available when operating polarized beams in the TS mode.

{\bf 1.}~Acceleration of the polarized beams.~---
Energy independence of the spin tune in figure-8 TS synchrotrons and racetracks with two snakes allows for acceleration of beams without polarization loss. Besides, the figure-8 design solves a serious problem of accelerating polarized $p$ and ${}^3 He$ beams in the booster energy range and polarized deuteron beams to the EIC energies, where full Siberian Snakes are technically challenging~\cite{b:AccJLEIC}.

{\bf 2.}~Long-term polarization maintenance.~---
Selection of optimal betatron tunes and compensation of the spin tune dependence on the energy spread allow one to reduce the depolarizing effect of higher-order nonlinear resonances thus enhancing the polarized beam lifetime~\cite{b:pol_RHIC}. 

{\bf 3.}~Polarization control.~---
An SN allows for flexible manipulation of the polarization direction not only at the IP but at any orbital location. It is also an efficient instrument for matching of the required polarization direction at injection of the beam into any of the synchrotrons of the collider complex. In addition, an SN simplifies the polarimetry by allowing adjustment of an optimal polarization orientation at a polarimeter.

{\bf 4.}~On-line monitoring of the polarization.~---
The TS mode gives a unique opportunity of quickly determining the polarization during an experiment, or on-line monitoring of the polarization. When adiabatically manipulating the polarization, its value is preserved at a high precision. The polarization vector is then a function of the SN magnetic fields and can be monitored using their set values. The relative accuracy $\Delta$ of the polarization direction obtained this way is determined by the ratio of the TS resonance strength to the SN strength: $\Delta\sim\omega/\nu_N \ll 1$.

{\bf 5.}~Spin flipping.~---
Spin flipping can be realized using a couple of SN solenoids that allow one 
to simultaneously control the polarization direction as well as the spin tune. 
The spin tune can be maintained constant during a spin flip thus avoiding any possibility of crossing the TS resonance or any of the higher-order spin resonances, which prevents polarization loss. To preserve the polarization degree during spin manipulation, the spin direction must change adiabatically. This condition can be specified as:
$\tau \gg 1/\nu$, where $\tau$ is the number of particle turns necessary to flip the spin. For JLEIC, the flip time requirements are 
\mbox{$t_p\gg 1$}~ms for protons and $t_d\gg 0.1$~s for deuterons~\cite{b:JLEICflip}.

{\bf 6.}~Ultra-high precision experiments.~---
It may be of interest to consider compensation of the coherent part of the TS resonance strength at a selected energy using an SN. Furthermore, an appropriate design of the magnetic lattice may significantly suppress the emittance-dependent part as well. This may create new opportunities for ultra-high precision experiments with polarized beams in synchrotrons such as search for a permanent electric dipole moment 
of charged spinning particles.

\textit{TS racetrack with two identical Siberian snakes.}~---
The TS mode of polarized colliding beams can also be realized in a synchrotron with a racetrack orbit geometry in its entire energy range by installing two identical Siberian Snakes in the opposite straights of the collider. For example, it is planned to use two solenoidal snakes in the NICA collider to operate it in the TS mode with polarized protons and deuterons in the momentum range of up to 13.5~GeV/c~\cite{b:NICA2}. 

\textit{TS mode in high-energy hadron colliders.}~---
The TS mode can also be efficiently used in high energy polarized $pp$ colliders such as RHIC, LHC and future ultra-high energy projects. Earlier studies have shown that, using a system of many snakes around a collider ring~\cite{Derbenev:1983tt,Derbenev:1988tt,b:SSC} or, alternatively, using spin-compensated quadrupoles, one could stabilize the coherent spin to hundreds of TeV and higher energies~\cite{b:spin_comp_quad}. Such systems can be adjusted to accommodate the TS spin dynamics with its advantages for spin control and manipulation discussed above. 

\textit{Proof-of-principle TS experiment in RHIC.}~---
RHIC is currently the only operating collider allowing experiments with polarized protons. 
In normal operation, the axes of its two helical snakes are orthogonal to each other and the spin tune is a half-integer~\cite{b:RHICSnakes}. The snake design allows one to change the snake axis orientations by adjusting the currents of helical magnet pairs. To experimentally test the TS mode, we plan to configure the snake axes to have the same orientations~\cite{b:testRHIC}.
Operation of RHIC in the TS mode may expand its capabilities for polarized proton experiments.

\textit{TS mode at an integer spin resonance.}~--- 
One may consider the possibility of arranging the TS mode at an integer spin resonance $\nu=\gamma G=k$ (an imperfection resonance) in a conventional racetrack synchrotron even without snakes. However, in contrast to the above examples, the energy dependence of the spin tune limits the efficacy of the TS mode at integer resonances. Nevertheless, integer resonances may be used to test and tune polarization manipulation devices (i.e. spin navigators) in the TS mode at low energies available at conventional synchrotrons such as COSY (Julich), AGS (Brookhaven), and Nuclotron (Dubna).
This mode of operation is also of interest for the EIC at BNL for manipulation of the deuteron polarization direction using spin navigators near the point of integer spin resonances, which occur about every 13~GeV. However, this question requires further study.

\textit{Conclusions.}~---
We presented a concept of a Transparent Spin method for control of hadron polarization in colliders and 
identified its main features. 
The Transparent Spin mode in figure-8 colliders and in racetracks with two identical snakes allows one to:
\begin{itemize}
    \setlength{\itemsep}{0pt}%
\item 
Control of the polarization by weak magnetic fields not affecting the orbital dynamics,
\item 
Accelerate the beam without polarization loss,
\item 	
Maintain stable polarization during an experiment,
\item	
Set any required polarization direction at any orbital location in a collider,
\item
Change the polarization direction using a spin navigator during an experiment,
\item
Monitor the polarization on-line during an experiment,
\item
Do frequent coherent spin flips of the beam to reduce experiment's systematic errors,
\item
Carry out ultra-high precision experiments.
\end{itemize}


The TS mode allows one to significantly expand the capabilities of polarized beam experiments at the EIC in the US~\cite{b:EIC_eRHIC}, NICA in Russia~\cite{b:NICA2}, EicC in China~\cite{b:EIC_China}, and other future facilities. It makes it possible for the RHIC-based EIC to manipulate the polarization during experiments in the whole energy ranges for protons and ${}^3He$ as well as for deuterons near energies corresponding to integer spin resonances. A proof-of-principle experimental test of the Transparent Spin mode in RHIC is currently in preparation~\cite{b:testRHIC}. The TS mode in the NICA collider with two solenoidal snakes allow for operation with polarized protons and deuterons in the whole momentum range of up to 13.5~GeV/c.

The TS concept advances the methodology of carrying out polarized beam experiments in fundamental physics and can stimulate corresponding detector development. It allows for high-precision measurements of the basic properties of a hadron such as its electric dipole moment.  

\begin{acknowledgments}
This material is based upon work supported by the U.S. Department of Energy, Office of Science, Office of Nuclear Physics under contract DE-AC05-06OR23177.
\end{acknowledgments}

\end{document}